**The Role of Interfacial Inherent Structures in the Fast Crystal Growth from Molten Salts and Metals**

Alexander Hawken, Gang Sun and Peter Harrowell

*School of Chemistry, University of Sydney, Sydney, New South Wales 2006, Australia*

Abstract

Molecular dynamics simulations of the temperature dependent crystal growth rates of the salts, NaCl and ZnS, from their melts are reported, along with those of a number of pure metals. The growth rate of NaCl and the FCC-forming metals show little evidence of activated control, while that of ZnS and Fe, a BCC forming metal, exhibit activation barriers similar to those observed for diffusion in the melt. Unlike ZnS and Fe, the interfacial inherent structures of NaCl and Cu and Ag are found to be crystalline. We calculate the median displacement between the interfacial liquid and crystalline states and show that this distance is smaller than the 'cage' length, demonstrating that crystal growth in the fast crystallizers can occur via local vibrations and so largely avoid the activated kinetics associated with the larger displacements associated with particle transport.

# 1. Introduction

The rate of crystal growth determines, along with crystal nucleation rates, a wide range of material properties that include microstructure [1], polymorph selection [2] and glass forming ability [3]. The propagation of a crystal interface into its supercooled melt also provides an experimentally accessible window onto the kinetics of arguably the most complex of structural fluctuations - the spontaneous reorganization of a liquid into a crystal. In the following, we shall refer to the *intrinsic* crystal growth rate V(T) defined as the steady state rate of crystal growth observable in the absence of kinetic control due to the diffusion of heat or impurities. This situation is most closely achieved experimentally at the growing tip of a crystal dendrite [4]. The intrinsic rate V(T) can be resolved into the product of a kinetic coefficient and a purely thermodynamic term [5], i.e.

$$V(T) = k(T)\left[1 - \exp\left(\Delta\mu / k_B T\right)\right] \qquad (1)$$

where $\Delta\mu = \mu_{xtl} - \mu_{liq}$ is the difference in chemical potential between the crystal and the melt and $k(T)$ represents the rate at which material is added to the crystal interface, measured as a length per unit time. The thermodynamic factor in Eq. 1 measures the difference, relative to $k$(T), between the rates of addition and subtraction of material at the interface.

The magnitude of the addition rate $k$(T) can vary enormously from material to material. The pure metals represent the fastest crystallizers with values of $k$(T$_m$) up to $10^2$ m/s [6] (where T$_m$ is the melting point) while SiO$_2$, one of the slowest crystallizers, has a value of $k$(T$_m$) ~ $10^{-8}$m/s [7]. Along with this range of 10 order of magnitude comes a striking variety in the temperature dependence of k(T). In most cases, crystal growth exhibits activated kinetics as characterized by an Arrhenius or super-Arrhenius temperature dependence of $k$(T) [8]. For many pure metals, however, the addition rate is not governed by an activation barrier even though the self diffusion in the associated melt is activated. This barrierless kinetics was noted by Coriell and Turnbull in 1982 [9]. They suggested that the rate of metal crystallization was determined by the speed of sound. This rate represents an upper bound on the rate at which the density change associated with crystal growth can propagate [10] but the authors of ref. [9] provided no physical explanation of how crystal growth avoids the constraints (responsible for activated behaviour) experienced by the same particle in the liquid. In the same year, Broughton et al [11] found that $k(T) \propto \sqrt{T/m}$ provided a reasonable fit to the crystal growth rate of the (110) and (100) faces of the Lennard-Jones face centred cubic (FCC) crystal into its melt. This rate, proportional to the mean thermal velocity of the atoms, is typically associated with the collision rate of a gas against a surface, not the dynamics of a supercooled liquid. In 2002, Jackson [12] concluded this fast crystal growth was the result of "a potential energy minimum for a liquid atom at the surface of a crystal". Recently [6], we have demonstrated that the local potential energy minimum of the liquid adjacent to the crystal interfaces in 6 FCC-forming metals is indeed crystalline, in clear contrast to the amorphous structure of the energy minima of the bulk liquid.

Our goal in this paper is to provide a quantitative explanation of this paradoxical behaviour in which crystal growth can occur without an activation barrier or, at least, with a barrier significantly smaller than that which governs particle diffusivity in the liquid. The structure of the crystal phase is a crucial component of this problem and so we would like to establish whether this accelerated growth can be observed in a category of materials with a larger diversity of crystal structures than those found in the pure metals. To do this we shall consider the crystal growth rates of ionic crystals from their melt.

Crystal growth kinetics of molten salts is important to a diverse range of phenomena including ceramic fabrication [13], phase change materials [14] and geological transformations [15]. Molten salts coolants - high heat capacity, low vapor pressure and chemical stability - have been recently proposed as heat carrying media in nuclear [16] and solar concentration [17] energy applications where the rates of crystal growth and melting are important design parameters [18]. While the growth rates of salts from solution has been extensively studied, studies of the crystal growth rates from the melt are limited. Okada and coworkers have reported on molecular dynamics simulations of the crystal growth rates NaCl [19], MgO [20] and CaCl$_2$ [21]. These preliminary studies indicate that the growth rates of the rock salt crystals – NaCl and MgO - are high, with maximum growth around 90 m/s reported. The maximum growth rate of CaCl$_2$ was ~ 2 orders of magnitude slower. In this paper we shall carry out detailed simulation studies of crystal growth of two canonical ionic crystals structures –the octahedral rock salt structure using NaCl, and the tetrahedral wurtzite structure exemplified by ZnS.

## 2. Algorithms and Models

The molecular dynamics simulations were carried out as follows. To accommodate the change in density on freezing, the system was maintained at zero external pressure by including free liquid surface on either side of the liquid and parallel to the crystal interface. To prevent temperature gradients the liquid was divided into layers 1nm thick and parallel to the crystal interface and the temperature was independently thermostated in each layer. The metals Cu, Ag and Fe are modelled using an Embedded Atom Model (EAM) potential due to Mendelev et al [22]. The NaCl is modelled using the Tosi-Fumi potential [23] and ZnS is modelled using a potential due to Grünwald et al [24]. The melting points are $T_m$ = 1275K (Cu), 1165K (Ag), 1775K (Fe), 1074K (NaCl) and 1750K (ZnS). The position of the crystal interface was monitored by the spatial distribution of order parameter (as defined in the Supplementary Material) and the growth rate V(T) is obtained as the slope of the linear evolution of the interface position, averaged over 10 runs. The (111) surface of the FCC crystals of Cu and Ag and the (110) surface of the BCC crystal of Fe are monitored (previous calculations found that no significant qualitative difference between the growth rates of different surfaces [6]) while for NaCl and ZnS, the (100) surface of the rock salt structure and wurzite structures were simulated. These simulations were carried out using the LAMMPS algorithms [25].

In our previous study of crystal growth in the pure metals [6], we noted that the metastability of the supercooled liquid may be bounded by a lower temperature limit $T_{sp}$. Below this temperature, crystallization occurs rapidly throughout the liquid, producing an effective kinetic spinodal. As a crystal growth rate is no longer meaningful once the liquid is unstable (i.e. $T < T_{sp}$) we need to determine these limiting temperatures. To do so, we cooled each liquid at a fixed rate measured the temperature at which crystal ordering is observed (see Supplementary Material). We find the liquid states of NaCl, Cu, Ag and Fe are effectively unstable below $T_{sp} = 600K$ ($0.56 \, T_m$), 852K ($0.67 \, T_m$) , 787K ($0.68 \, T_m$) and 1078K ($0.61 \, T_m$), respectively. ZnS, in contrast, exhibits no sign of an instability via the protocol used here.

## 3. Crystal Growth Rates and the Magnitude of the Associated Activation Energy

The calculated growth velocities for NaCl, ZnS, Cu, Ag and Fe are plotted against $T/T_m$ in Fig.1 down to their respective values of $T_{sp}$. We find that NaCl crystals exhibit extremely high growth rates, similar in both magnitude and temperature dependence to the metal growth rates. These rates are remarkable given the compositional ordering that must accompany crystal growth in the salt.  The salt ZnS and the BCC metal Fe present clear contrasts in crystal growth behaviour to that of these fast crystallizers. The growth rates of ZnS and Fe exhibit a marked turnover and slow down at large supercoolings, quite different from the monotonic increase in growth rate of NaCl and the FCC metals. In the case of ZnS, this difference extends to the magnitude of the growth rate with ZnS exhibiting significantly slower growth. The Fe crystal, in contrast, is found to grow faster than Cu.

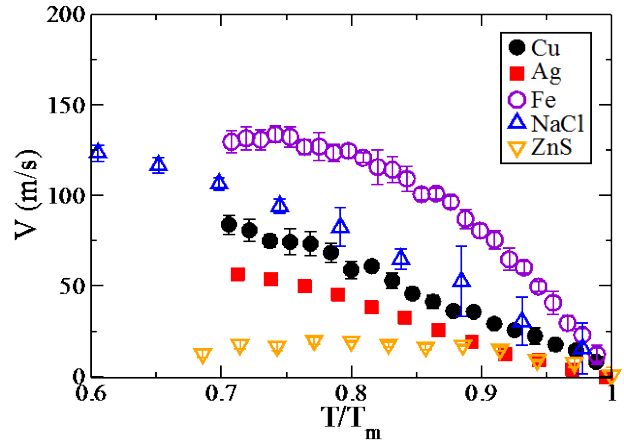

**Figure 1.** A plot of the crystal growth rate V(T) against T/T$_m$ of NaCl, ZnS, Cu, Ag and Fe. Tabulated growth rates are provided in the Supplementary Material.

From the growth rate V(T) we can extract the rate of addition $k$(T) using Eq. 1 and the approximation $\Delta\mu = \frac{\Delta h(T_m)}{T_m}(T - T_m)$. The enthalpy change on fusion per atom $\Delta h(T_m)$ was obtained simulation of liquid and crystal at T$_m$  $\Delta h(T_m)$ = 2.13(Cu), 1.77 (Ag), 2.53 (Fe), 2.43 (NaCl) and 3.10 (ZnS) x 10$^{-20}$J. The resulting addition rates are plotted in Fig.2a against 1/T. We find that the T dependence of $k$(T) can be reasonably treated as Arrhenius, i.e. $\ln k = c - E_a / k_B T$ . The values of each activation energy E$_a$ obtained by this fit are presented in Table 1. The activation energy E$_D$ of the self diffusion coefficient D in the melt has also been determined in Fig. 2b and presented in Table 1. We find that in ZnS and Fe the activation energies of crystal addition and diffusion are similar, a result consistent with the 'classical' Wilson-Frenkel expression for k(T) [26]. In clear contrast, the activation energies for k(T) in NaCl and the FCC metals are significantly smaller than that of their diffusion coefficients. The explanation of this difference between crystal addition and diffusion is the subject of the remainder of this paper.

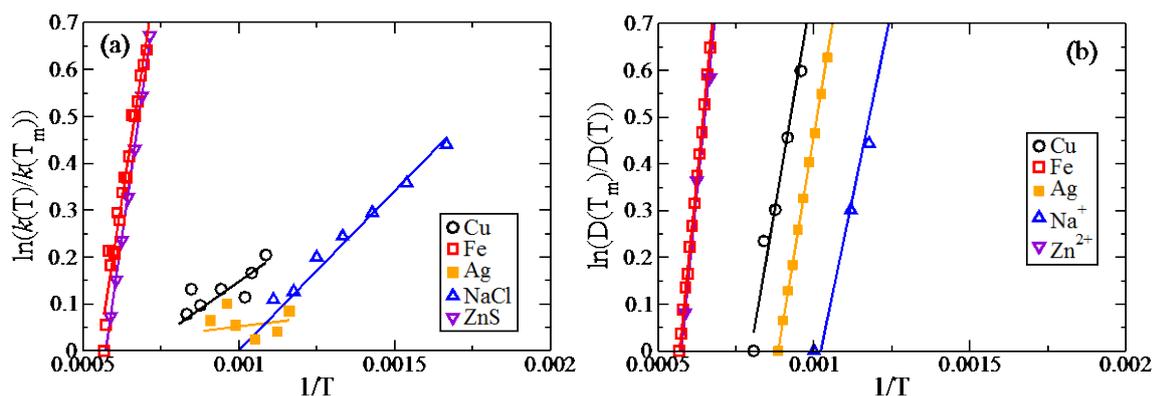

**Figure 2.** a) The plot of ln[k(T)/ k(T_m)] vs 1/T for NaCl, ZnS, Cu, Ag and Fe. b) The plot of ln (D(T_m)/D(T)) vs 1/T for Na⁺, Cl⁻, Zn², S²⁻ , Cu, Ag and Fe. (The cations have the larger diffusion coefficient in each salt.)  In each case the slope of the fitted line is equal to the associated activation energy, $E_a$ and $E_D$, for k(T) and D(T), respectively.

|  | Cu | Ag | NaCl | ZnS | Fe |
|---|---|---|---|---|---|
| $E_a$ | 463.6 | 81.7 | 681.5 | 4686.8 | 4153 |
| $E_D$ | 3900.6 | 4023.7 | 3196.5 (Na⁺) | 6570 (Zn²⁺) | 6574 |

**Table 1.** The activation energies $E_a$ and $E_D$ for k(T) and D(T) for Cu, Ag, NaCl, ZnS and Fe as obtained from the slopes of the fitted lines in Fig. 2. For the salts we consider the cation diffusion coefficient as this is larger than the anion in each case.

## 4. Interfacial Inherent Structure and the Proximity of Crystal and Liquid

The near absence of an activation barrier to crystal growth in pure metals and NaCl suggests that the liquid does not need to cross an energy barrier to organise itself into a crystal. The interfacial liquid and crystal, in other words, share a common potential energy basin. In 1982, Stillinger and Weber [27] suggested that a liquid structure might be usefully regarded as vibrational perturbations around structures corresponding to local minima of the potential energy. These so-called 'inherent structures' have since become widely used in the normal mode treatment of liquid dynamics [28], the determination of configurational entropy [29] and in master equation models of relaxation dynamics [30]. In 3D, the inherent structures of liquids are typically disordered. In the case of ionic liquids, inherent structures for $ZnBr_2$ [31], $MgF_2$ [32] and mixtures of NaCl-CsCl [33] have been reported without any mention of crystalline structures. We have examined the inherent structures of the bulk melts of NaCl and ZnS. The fraction of crystal-like structures in the NaCl inherent structures formed from the 1070K liquid is less than 7%. When crystallinity was detected it occurred as well defined clusters. The inherent structures of the ZnS melt exhibited little in the way of crystallinity and no sign of the local compact clusters observed in NaCl. Details of these calculations are provided in the Supplementary Material.

What of the inherent structures of the liquid directly adjacent to the crystal interface? In ref. 6, we demonstrated that the inherent structure of the liquid adjacent to the crystal-liquid interface of pure metals is crystalline. This is evident by the observed propagation of the interface into the melt during a conjugate gradient minimization of the potential energy. This observation is significant in the context of barrierless growth kinetics since the minimization specifically excludes the possibility of crossing energy barriers. In the two salts studied in this paper, we have two quite different temperature dependencies of crystal growth rates and so can put to the test the proposition of ref. 6 that the small activation energies are caused by crystalline interfacial inherent structures. As shown in Fig. 3, the NaCl interface advances significantly during minimization. In contrast, the ZnS and Fe interfaces do not move at all, as shown in Figs. 4 and 5. That the presence or absence of crystallinity in interfacial inherent structures correlates with barrierless or activated growth kinetics, respectively, provides support for our previous conclusion [6] that it is the crystalline order of the interfacial inherent structure that is responsible for the low activation energy of the crystal growth. The distance L of propagation of the interface during minimization (indicated in Fig. 3) provides a useful measure of the structural susceptibility of the liquid to the crystal interface. The values of L, averaged over 10 configurations (i.e. 20 interfaces), for the liquids studied in this paper and those considered in ref. 6 are reported in Table 2.

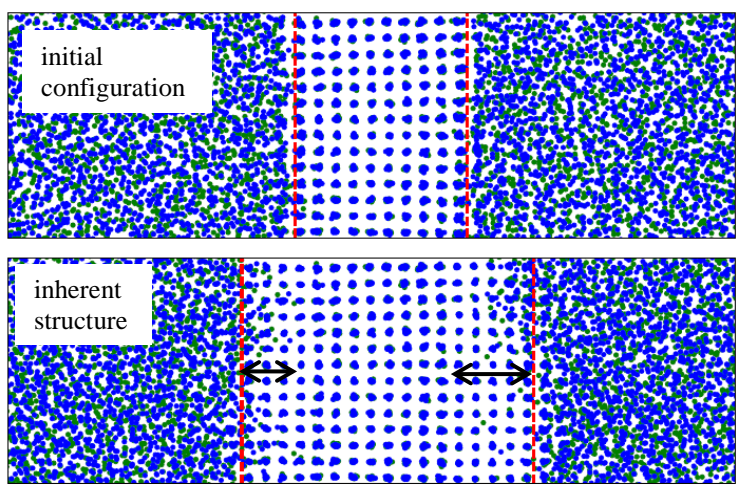

**Figure 3.** Growth of the rock salt (100) interface for NaCl during the energy minimization of the interface shown by comparison of the initial (top) and inherent structure (bottom) configurations. The crystal interface in each case is indicted by the red line. The interfacial growth is indicated by the black arrows. Averaged over the minimization of four distinct configurations, the average growth distance L = 8.44 Å

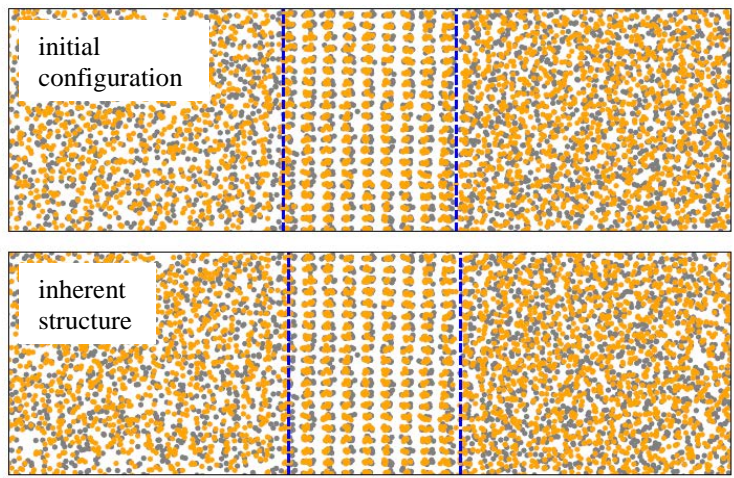

**Figure 4.** The change in the wurzite (100) interface for ZnS during the energy minimization of the interface shown by comparison of the initial (top) and final (bottom) configurations. The crystal interface in each case is indicted by the blue line. No significant movement of the interface was noted.

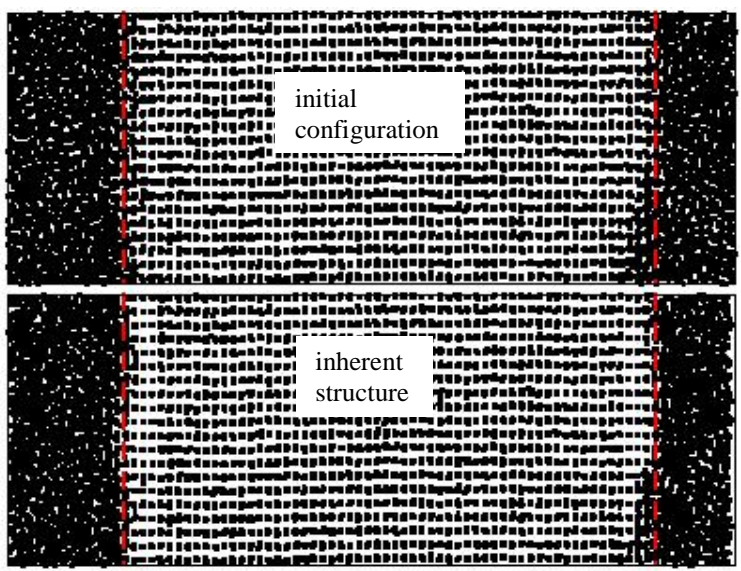

**Figure 5.** The change in the BCC (110) interface for Fe during the energy minimization of the interface shown by comparison of the initial (top) and final (bottom) configurations. The crystal interface in each case is indicted by the red line. No significant movement of the interface was noted.

To go beyond this qualitative association, we must quantify the proximity of crystal to liquid identified by the energy minimization of the crystal-liquid interface. The distribution of atomic displacements associated with the movement of each atom from its position in the initial configuration to that in the crystal inherent structure as a result of energy minimization provides this proximity measure. Note that we include only the displacements of those particles initially liquid that were mapped to a crystalline site by energy minimization. In Fig. 6 we plot the distribution of displacements for the NaCl and Cu interfaces. We have no matching set of ordering displacements for ZnS and Fe since their minimization did not result in a map to the crystalline state. This means that the following analysis cannot be extended to these systems until we can develop an analogous definition of ordering displacements that does not depend on the existence of crystalline interfacial inherent structures. Each displacement has been resolved to the in-plane component (i.e. the component of the displacement in the xy plane parallel to the interfacial plane) and the component along the interfacial normal (here parallel to the z axes). As shown in Fig.6a, the normal component $d_z$ for NaCl is significantly larger than the in-plane component $d_{xy}$. This difference arises because minimization includes a general densification of the liquid phase, a change characterised by the collective movement of the liquid particles towards the crystal interface which contributes an additional component to atomic displacements in the normal direction. The large normal displacement $d_z$ in NaCl reflects the large (~ 70%) density increase on freezing. Cu, like most of the pure metals, exhibits only a small density increase at freezing (~ 4%) and, consequently, the normal displacement is not very different from that in the x or y direction.

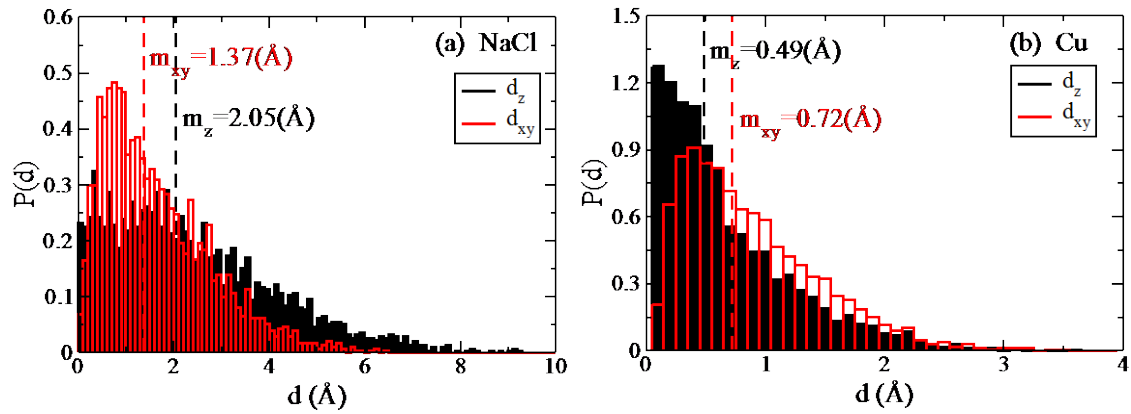

**Figure 6.** Distribution of the in-plane ($d_{xy}$) and normal ($d_z$) atomic displacements arising from the minimization of the energy in the interfacial regions of a) NaCl and b) Cu. The median values $m_{xy}$ and $m_z$ are indicated by vertical dashed lines.

| lengths (Å) | NaCl | Al | Cu | Ni | Ag | Pt | Pb |
|---|---|---|---|---|---|---|---|
| L | 8.44 | 15.3 | 6 | 4 | 5 | 8 | 6 |
| m | 1.68 | 1.2 | 0.88 | 0.92 | 0.83 | 0.94 | 0.92 |
| $r_1$ | 2.59 | 2.75 | 2.45 | 2.45 | 2.85 | 2.75 | 3.45 |
| lattice spacing | 3.23 | 2.3 | 2.09 | 2.03 | 2.36 | 2.26 | 2.86 |

**Table 2.** Values of L for the interfacial inherent structure (see text), the median in-plane displacement m (defined below), $r_1$, the (scaled) position of the 1$^{st}$ peak of the liquid radial distribution function at $T_m$ and the crystal lattice spacing normal to the interface, for NaCl, Cu, Ag and the other four metals considered in ref. 6. Note the anomalously large value of L for aluminium.

## 5. The Reorganization Time and the Rate of Crystal Growth

We are interested in linking a time scale to a characteristic displacement. The mean displacement is dominated by the long tail in the displacement distribution. As the particles associated with these large displacements are few in number, they are unlikely to play a dominant role in the general development of order. Instead, we shall use the median displacement $m$ to characterise the distribution. To assign a time scale to the length $m$ we shall resort to the simple expedient of comparing the distance $m$ to the time dependent mean squared displacement of the bulk liquid at the appropriate temperature. This proposal assumes that, having identified the ordering length from the inherent structure calculation, the incoherent single particle motion, as described by the liquid mean squared displacement, represents a reasonable representation of the type of motion involved. This argument does not apply to the collective compression that contributed to the normal displacement, $m_z$ as in the case of NaCl, and so we shall use a length $m = m_{xy} \times \sqrt{3/2}$, scaling the in-plane median displacement to be an effective 3D displacement in all cases. Values of $m$ for NaCl and the metals: Pb, Cu, Ni, Pt, Ag and Al, are provided in Table 2. Some sense of the relative magnitude of these ordering lengths can be reached by comparing this length with that corresponding to the first peak of the radial distribution function in the liquid $r_1$ (a length similar to the lattice spacing in simple crystal phases). The values of $r_1$ for T = $T_m$ are also included in Table 2. We note that $m < r_1$ in all the systems studied, a point we shall return to below.

A temperature dependent time scale is assigned to the squared median displacement $m^2$ using the liquid mean squared displacement as shown in Fig. 7. While we consider $m$ to be effectively T independent, we note that the associated time scale will exhibit a T dependence due to the dependence of the liquid mobility on temperature. A value of *m* that lies in the short time ballistic regime will exhibit a weak temperature dependence $\sim \sqrt{T}$ while a small increase in *m* will see it lie in the diffusive regime and so characterised by an activated T dependence. The analysis shown in Fig. 7 highlights a central result of this paper – that the proximity of the crystal and liquid, as measured by *m*, plays the crucial role in determining the character of the T dependence of the growth rate. Highlighting this, is the observation that the movement over a distance equal to the nearest neighbour distance $r_1$ in the liquid, typically $\sim$ 2-4 *m*, always lies in the region of activated kinetics.

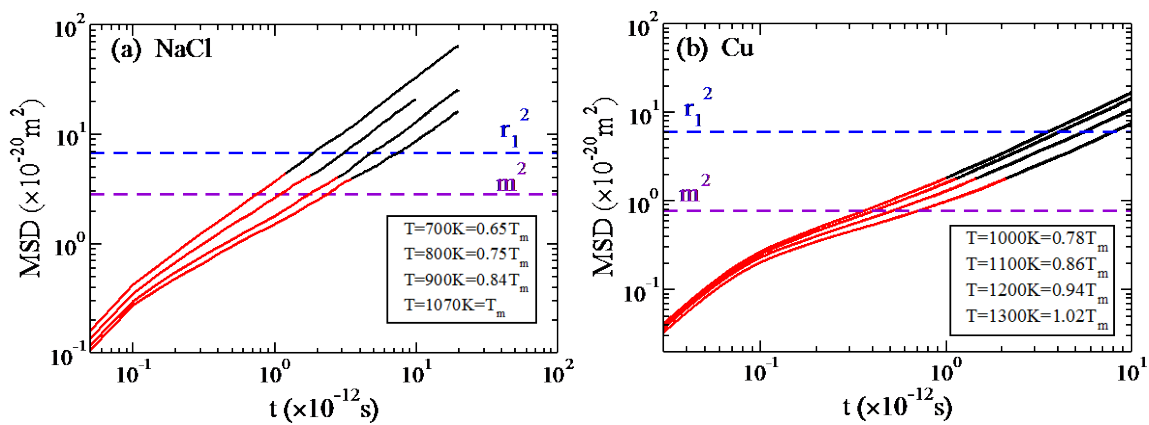

**Figure 7.** The mean squared displacements (MSD) vs time for a) the Cl⁻ ions in NaCl, and b) Cu for the indicated values of T (the higher curves corresponding to higher temperatures). The different curves correspond to different temperatures with the larger value of MSD associated with a larger value of T (as indicated). The diffusive regime on each curve is indicated in black and the non-diffusive regime (ballistic and crossover) is indicated in red. The values of $m^2$ and $r_1^2$ are indicated by horizontal lines and the associated time $\tau$ is read off from the intersection of this line with that of the mean squared displacement at each value of T.

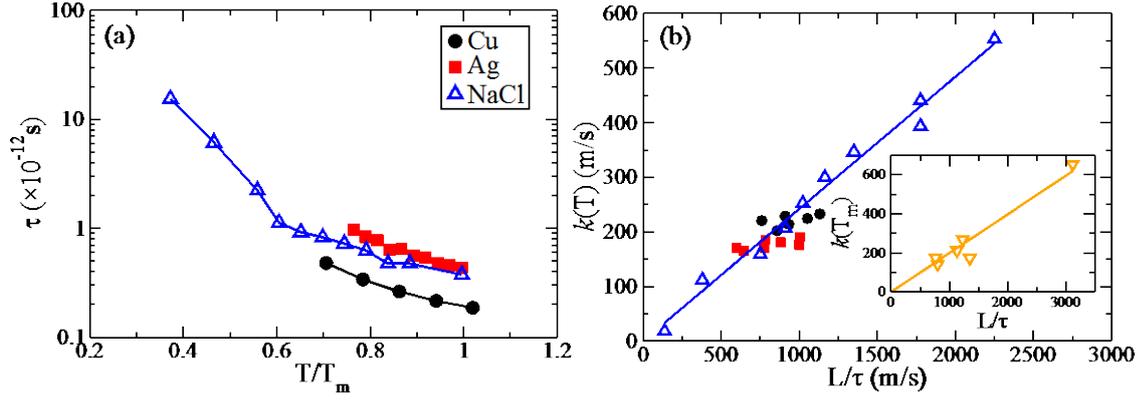

**Figure 8.** a) The reorganization time $\tau$ vs T as obtained from the construction in Fig. 7 for NaCl, Cu and Ag. b) Plot k vs $L/\tau$ for NaCl, Cu and Ag. Insert. Plot of $k(T_m)$ vs $L/\tau(T_m)$ for the six metals. The lines in both plots represent the fit $k(T) = cL\tau^{-1}(T)$ with c =0.22. The value of L for each system is reported in Table 2.

The temperature dependence of the reorganization time $\tau$, as plotted in Fig. 8a, arises from the T dependence of the mean squared displacements (see Fig. 7) used to convert $m$ to $\tau$. For the metals Cu and Ag, $\tau$ changes little over the temperature range of interest. In the case of NaCl, we see a crossover in behaviour, with a similarly small temperature dependence of $\tau$ over most of the observed range but a clear switch to a much stronger T dependence at the lowest supercooling. This is a result of the decrease in the caging distance below $m$ in the salt at these lower temperatures.

Having obtained the characteristic time $\tau(T)$ from the displacement distributions and the liquid mobilities, we can now test a simple model of the crystal addition rate, i.e.

$$k(T) = c \times L / \tau(T) \qquad (2)$$

where we have included both the estimate of the time of transformation from liquid to crystal and the length L, that characterises the extent of the crystal domain generated during energy minimization. In Fig. 8b we plot k(T) vs L/τ and find that the linear relationship proposed in Eq. 2 provides a good representation of the crystal growth data. Furthermore, the anomalously large value of k(T) found previously [6] for Al can now be accounted for (see Fig. 8b insert) in terms of the particularly large value of L for this metal. We find a single value of c = 0.22 provides a good fit for the data for all the metals and for NaCl.

## 6. Conclusions

We have demonstrated that the existence of a crystalline interfacial inherent structure is correlated with small activation energies for crystal addition during growth, where 'small' means values of $E_a < 0.25\ E_D$, and $E_D$ is the activation energy for liquid diffusion. This correlation applies to both crystal growth of pure metals and ionic crystals from the melt. ZnS and Fe crystallization, which both show no sign if crystalline order in its interfacial inherent structure, exhibits a large activation energy similar in magnitude to that of liquid diffusion. The distribution of displacements associated with minimization provides a quantitative measure, $m$, of the proximity of the interfacial liquid and the crystal. The time scale τ associated with movement over this distance is shown to provide a quantitative account of the addition rate k(T) with k(T) = cL/τ. The length L, the thickness of the crystalline component of the interfacial inherent structure, represents a measure of the structural susceptibility of the liquid to the adjacent crystal.

Where previous theories of crystal growth have regarded the liquid as a continuum and an unstructured source of atoms for crystal addition, the treatment presented here treats the disordered liquid configurations explicitly. Only by doing so is it possible to obtain the proximity length $m$ – the magnitude of which plays an essential role in determining the temperature dependence of the associated crystal growth rate. This distinction is crucial. In the density wave theory of crystal growth due to Mikheev and Chernov [34], for example, the time scale is set by the relaxation time for density fluctuations in the liquid over the length scale of the crystal layer spacing, i.e. ~ $r_1$. As shown in Fig. 6, this length is large enough to always correspond to control by activated kinetics.

We have established that the effectively barrierless crystal growth kinetics in NaCl and the FCC-forming metals is a consequence of the fact that the liquid adjacent to the crystal is sufficiently proximate to the crystal structure that only displacements on the scale of vibrational amplitudes are required to transform the liquid to crystal. The corollary of this conclusion would be that the strongly activated growth observed in ZnS and Fe is a consequence of *larger* ordering displacements. We cannot confirm this until we have developed an alternative definition of the proximity length *m* that does not require the interfacial inherent structure to be crystalline, i.e. the situation for ZnS and Fe. Work in progress is focussed on developing this alternative definition. Such alternate measures of structural proximity will permit the extension of the approach to explore the nature of activated control of addition during crystal growth.

**Acknowledgements**

We gratefully acknowledge the support of the Australian Research Council and computational support from the Computational Chemistry Facility of the School of Chemistry and the High Performance Computing Facility of the University of Sydney.

**Supplementary Information**

**The Role of Interfacial Inherent Structures in the Fast Crystal Growth from Molten Salts and Metals**

Alexander Hawken, Gang Sun and Peter Harrowell

**Contents**

## 1. The Identification of the Limit of Metastability $T_{sp}$

The limit of stability of a supercooled liquid is defined as the temperature at which the liquid freezes while being cooled at a rapid rate.( Previously [1] we have shown that the value of $T_{sp}$ is only weakly dependent on the cooling rate, at least over a modest range.) Freezing, here, is identified by a rapid drop in the potential energy U. In the results plotted in Fig. 1, NaCl, ZnS, Cu and Ag, were cooled at rates of 1.9K/ps, 1.9K/ps, 0.25K/ps and 0.125K/ps, respectively.

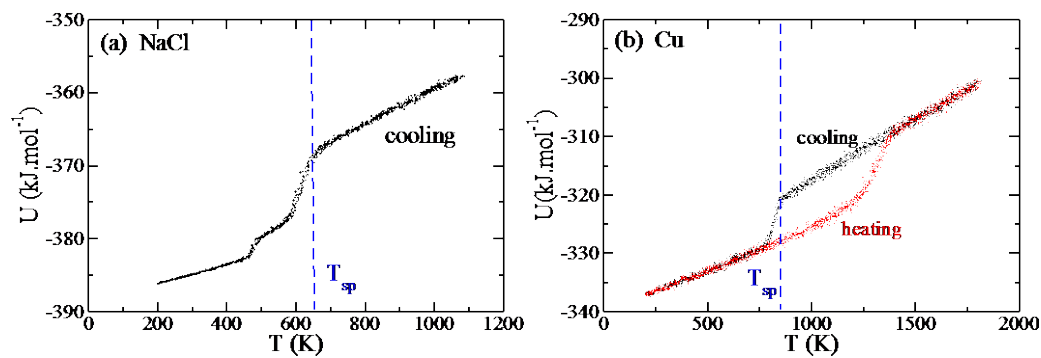

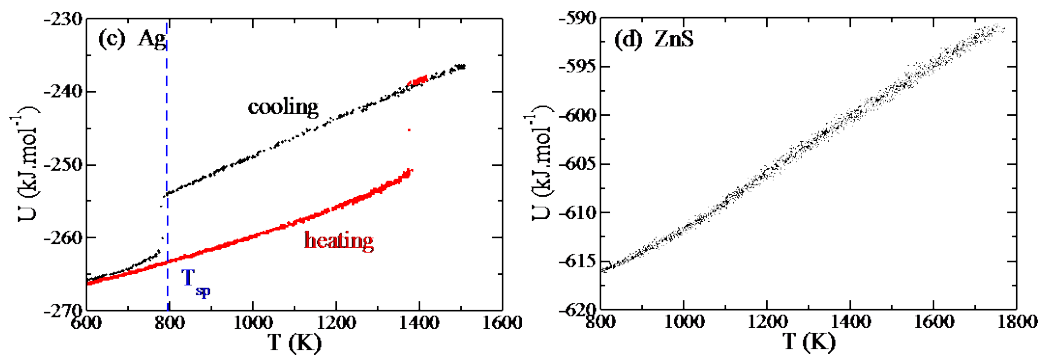

**Figure S1**. Plots of potential energy U versus T for constant cooling rate quenches of the homogeneous melts (a) NaCl, (b) Cu, (c) Ag and d) ZnS. The abrupt drop in U in a), b) and c) identify the effective spinodal temperature $T_{sp}$ . The absence of any abrupt drop in U in the ZnS melt indicates that this melt shows no sign of an instability.

The energy U vs T has been determined for both a cooling run starting at the melt and a heating run starting from the final configuration of the cooling run. The observed hysteresis between cooling and heating confirms that the drop in potential energy is the result of a phase transition. The values of $T_{sp}$, as determined from Fig. 1, are 852K (Cu), 787K (Ag) and 600K (NaCl). Using the same approach, a value of 1078K was found for the $T_{sp}$ of Fe.

## 2. The crystal growth rates of Cu, Ag, NaCl and ZnS down to T = $T_{sp}$

| Cu | | Ag | | Fe | | NaCl | | ZnS | |
|---|---|---|---|---|---|---|---|---|---|
| T (K) | V(m/s) | T (K) | V(m/s) | T (K) | V(m/s) | T (K) | V(m/s) | T (K) | V(m/s) |
| 1275 | 0 | 1165 | 0 | 1775 | 0 | 1074 | 0 | 1750 | 0 |
| 1260 | 8.0944 | 1130 | 3.965 | 1760 | 12.011 | 1050 | 15.475 | 1700 | 7.554 |
| 1240 | 14.311 | 1100 | 8.895 | 1740 | 22.86 | 1000 | 30.347 | 1650 | 8.979 |
| 1220 | 17.688 | 1070 | 12.265 | 1720 | 29.43 | 950 | 52.488 | 1600 | 14.977 |
| 1200 | 22.319 | 1040 | 18.909 | 1700 | 40.67 | 900 | 64.832 | 1550 | 17.211 |
| 1180 | 25.547 | 1010 | 25.406 | 1680 | 49.63 | 850 | 82.44 | 1500 | 16.187 |
| 1160 | 29.083 | 980 | 32.495 | 1660 | 60.14 | 800 | 93.992 | 1450 | 17.764 |
| 1140 | 35.682 | 950 | 38.085 | 1640 | 64.67 | 750 | 106.362 | 1400 | 19.458 |
| 1120 | 36.1 | 920 | 45.275 | 1620 | 75.4 | 700 | 116.484 | 1350 | 19.849 |
| 1100 | 41.321 | 890 | 49.735 | 1600 | 80.5 | 650 | 123.351 | 1300 | 16.565 |
| 1080 | 45.772 | 860 | 53.6 | 1580 | 87 | | | 1250 | 17.435 |
| 1060 | 52.871 | 830 | 56.456 | 1560 | 96.3 | | | 1200 | 12.433 |
| 1040 | 60.925 | | | 1540 | 101 | | | | |
| 1020 | 58.867 | | | 1520 | 100.6 | | | | |
| 1000 | 68.294 | | | 1500 | 109.18 | | | | |
| 980 | 73.265 | | | 1460 | 115.7 | | | | |
| 960 | 74.2889 | | | 1400 | 123.6 | | | | |
| 940 | 74.9333 | | | 1300 | 130.8 | | | | |
| 920 | 80.6556 | | | 1200 | 133.8 | | | | |
| 900 | 83.855 | | | 1100 | 121.6 | | | | |

**Table S1.** Steady state crystal growth rates for Cu, Ag, NaCl and ZnS as a function of temperature. The melting points $T_m$ are indicated for each material by the V = 0m/s entry.

## 3. Inherent structure of molten salts

The inherent structures of the molten salts were generated by minimzing the potential energy of an instantaneous configuration of the liquid. Conjugate gradient energy minimizations using the Polack-Ribière algorithm were performed on the instantaneous configurations, with a relative energy ratio of less than $1 \times 10^{-8}$ between steps as the only stopping criterion. When an iteration resulted in a decrease in energy between steps of less than this ratio, the quench was considered finished and particle coordinates recorded.

The degree of crystalline order in the inherent structures was monitored by two methods: calculation of the radial distribution functions and calculation of the crystalline order parameters as described in the text. We shall start by considering the NaCl system.

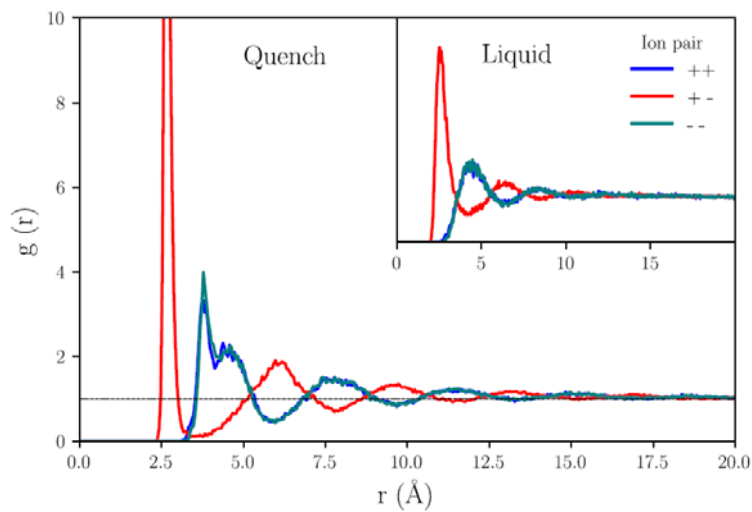

**Figure S2.** Pair-wise distribution function for the molten NaCl system before (insert) and after quenching from 2000K. The first peak of the pair correlation between like ions shows signs of a splitting in the inherent structure.

Inspection of the change in the pair distribution functions in Fig. S2 shows little average change beyond the narrowing of the peaks with the absence of thermal vibrations. Some new structure does appear in the first peak of the distribution of separations between like ions. The degree of crystalline order for the inherent structures of the molten salts can be considered by analysing the local orientational order parameter of each ion. For NaCl inherent structures, 7% of the ions are in crystalline environments with some tendency to cluster. The largest crystalline cluster, shown in Fig. S3, was formed of 31 ions. The average potential energy of each ion in the inherent structure was 20 kJ.mol⁻¹ lower than that of the molten liquid at 1070 K for NaCl. For ZnS the average potential energy of ions was lowered by 10 kJ.mol⁻¹ from 2500 K upon quenching, and no crystalline order was developed.

This process was repeated for ZnS, with 16848 ions being melted from their wurtzite lattice at 2500 K before being cooled and quenched in an identical manner to NaCl. The first peak in the cation-cation correlation function clearly splits while that of the anion-anion correlation shows no sign of splitting.

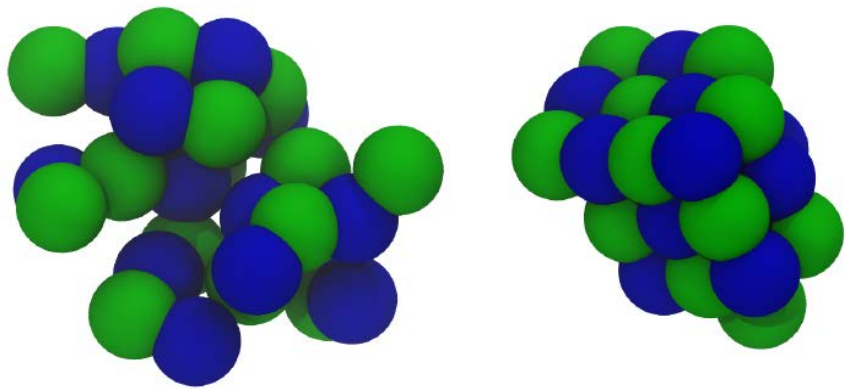

**Figure S3.** The initial (left) and inherent (right) structures of region in a NaCl melt that exhibits crystal order following a conjugate gradient minimization from the melt at 1070K.

This break in the symmetry of like ion correlation, seen in NaCl, is consistent with the greater size discrepancy between the ions. There is a sign of some additional structure in the 2^nd peak of the correlation between unlike ions. Trying to separate out the crystalline regions in ZnS inherent structures proved much less conclusive than in the case of NaCl. The number of particles identified as crystalline was low and these particles showed little sign of clustering and we conclude, despite the changes in the correlation function, that whatever structure does develop in the ZnS inherent structures, it cannot be classified as wurtzite-like.

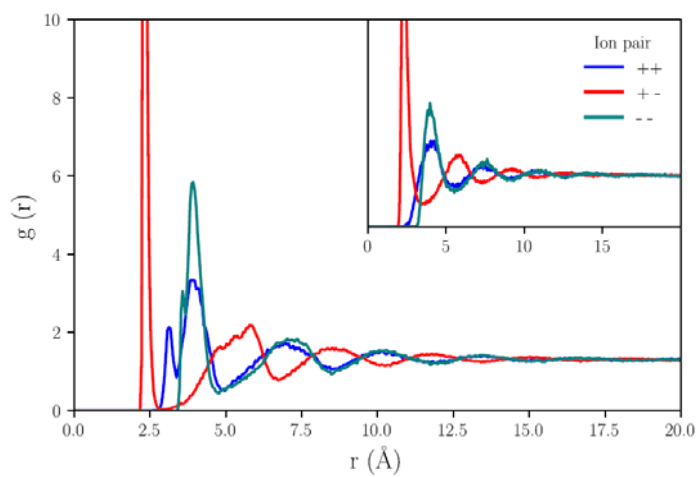

**Figure S4.** Pair-wise distribution function for the molten ZnS system before (insert) and after quenching. The first peak of the pair correlation between cations shows a clear splitting in the inherent structure.